\documentstyle[aps,preprint]{revtex}
\input epsf
\global\firstfigfalse  

\tighten

\def\hh{\cr\noalign{\vskip2truemm}}

\def\pn{${\mathcal I}^-$}
\def\fn{${\mathcal I}^+$}
\def\Hin{${\mathcal{H}}_{\rm in}$}
\def\fin{\phi_{\rm in}}
\def\bfin{\overline{\phi}_{\rm in}}
\def\Hout{${\mathcal{H}}_{\rm out}$}
\def\fout{\phi_{\rm out}}

\begin{document}
\pagestyle{empty}

\preprint{
\begin{minipage}[t]{3in}
\end{minipage}
}

\title{
Black hole evaporation based upon a $q$-deformation description }

\author{Xin Zhang \\\bigskip}
\address{Institute of High Energy Physics, Chinese Academy of Sciences,\\
P.O.Box 918(4), Beijing 100049, China\\
\smallskip{\tt zhangxin@mail.ihep.ac.cn}}

\maketitle

\begin{abstract}

A toy model based upon the $q$-deformation description for
studying the radiation spectrum of black hole is proposed. The
starting point is to make an attempt to consider the spacetime
noncommutativity in the vicinity of black hole horizon. We use a
trick that all the spacetime noncommutative effects are ascribed
to the modification of the behavior of the radiation field of
black hole and a kind of $q$-deformed degrees of freedom are
postulated to mimic the radiation particles that live on the
noncommutative spacetime, meanwhile the background metric is
preserved as usual. We calculate the radiation spectrum of
Schwarzschild black hole in this framework. The new distribution
deviates from the standard thermal spectrum evidently. The result
indicates that some correlation effect will be introduced to the
system if the noncommutativity is taken into account. In addition,
an infrared cut-off of the spectrum is the prediction of the
model.

\vskip .4cm


\vskip .2cm


\end{abstract}

\newpage
\pagestyle{plain} \narrowtext \baselineskip=18pt

\setcounter{footnote}{0}

It is well known that the noncommutative geometry \cite{connes1}
plays a very important role in revealing the properties of Planck
scale physics \cite{connes2,seiberg,yoneya}. It has been suspected
for a long time that the noncommutative spacetime might be a
realistic picture of how spacetime behaves near the Planck scale.
The most realistic laboratory for testing Planck scale physics is
the early universe and the black hole. In the studies of the early
universe, it has been shown \cite{inflation} that if the spacetime
is indeed noncommutative on short distance, this may have an
impact on early universe such as the evolution of the primordial
fluctuation spectrum generated during the inflation (for spacetime
noncommutative inflation, see also \cite{li}). Moreover, it can be
imagined naturally that the spacetime noncommutative effects in
the vicinity of black hole horizon might bring influence to the
fundamental physics in some degree. However, the investigation for
testing those noncommutative effects near black hole is extremely
difficult due to the fact that a consistent theory of quantum
gravity is not available at present, let alone a theory of
noncommutative quantum gravity. Still, some efforts have been made
on this subject \cite{unruh,yan,v}. It should be mentioned that by
considering a semi-classical theory of noncommutative fields, the
authors of Ref.\cite{yan} give some remarks on 't Hooft's brick
wall model \cite{thooft}.

In this note, we will make an attempt to consider the spacetime
noncommutative quantum effects near the black hole by using a
deformation description of quantum fields. The deformation
description of physical systems can be traced back to the quantum
group symmetry. Quantum groups lead to an algebraic structure that
can be realized on quantum spaces, and such quantum spaces can be
interpreted as noncommutative configuration spaces for physical
systems \cite{wess}. Moreover, it has been demonstrated in Ref.
\cite{wess} that in the $q$-deformed algebraic structure the
spacetime occurs in different phases --- a lattice phase and a
continuous phase. It might be that for very small distances and
high energy density the lattice phase is dominant and provides a
natural ultraviolet cutoff. Studies on the deformation description
for physical systems have made great progresses. The dynamics of a
deformed photonic field interacting with matter presented many
interesting properties in the area of the quantum optics
\cite{optics}. Great success has also been achieved by making use
of a deformed oscillator to study the molecular spectrum
\cite{molecular}. In addition, the thermodynamics of a deformed
oscillator system has been investigated deeply \cite{gas}. In
particular, it should be pointed out that the $q$-deformed
noncommutative theory has been presented for resolving the origin
of the ultrahigh energy cosmic ray and the TeV-photon paradoxes
\cite{gzk}.

The prediction of black hole evaporation \cite{hawking} is a great
triumphant of the combination of general relativity and quantum
mechanics, even though this combination is not complete. However,
the laws of quantum fields in curved spacetime are only
semi-classical such that which should be valid only for cases of
relatively low energy scale. As the size of black hole approaches
extremely small distance scale, for instance, the Planck scale or
so, the spacetime noncommutativity might play an important role in
some courses of fundamental physics, e.g. the black hole
evaporation. The consideration of spacetime noncommutativity
perhaps helps resolving the problem of information loss
\cite{information} (if the radiation spectrum deviates the
standard thermal spectrum, it seems that some quantum hair
appears). Though we still have no enough knowledge about the
underlying theory --- the quantum gravity, we can still make an
attempt to consider some noncommutative effects in the framework
of quantum field theory in curved spacetime. Therefore, we propose
a toy model for testing this noncommutative effect in the process
of black hole evaporation. The toy model is based upon such an
assumption that from an effective point of view the behavior of
the spacetime noncommutative effects is reflected completely by
the deformation of the radiation field meanwhile the background
metric is preserved as usual. It has been found that such a simple
assumption will give rise to interesting results. First, the
spectrum we obtained deviates from the perfect thermal spectrum
evidently, which indicates that some correlation effects are
introduced apparently. This maybe helpful for understanding the
problem of information loss. Second, the new spectrum takes on
some weird behavior, i.e. infrared divergency. It is interesting
for us, because that divergency always implies new physics, so we
can view that some unknown physical mechanism which we neglect
plays the role. On the other hand, however, perhaps our method is
not suitable for low frequencies. In any case, an infrared cut-off
is necessary, which seems to take on some UV/IR correspondence
principle (e.g. see \cite{uvir}), since the infrared cut-off is
the consequence of the ultraviolet modification \cite{wess}.

In what follows we will introduce the deformation description to
the calculation of the radiation spectrum of the Schwarzschild
black hole evaporation. For simplicity only the bosonic field is
taken into account in this note. The natural quantum of the
quantum field theory on noncommutative geometry is the so-called
deformed harmonic oscillator. On the topic of deformed oscillator,
a lot of papers have been published \cite{oscillator}. Similar to
the case of an ordinary quantum field system, the properties of a
noncommutative quantum field system can be described by a
collection of the deformed harmonic oscillators with the creation
and annihilation operators satisfying the following $q$-deformed
Weyl-Heisenberg algebraic relations
\begin{equation}
a_qa_q^\dagger-qa_q^\dagger
a_q=q^{-N},~~~~[N,a_q]=-a_q,~~~~[N,a_q^\dagger]=a_q^\dagger.\label{q}\end{equation}
Obviously, the deformed Weyl-Heisenberg algebra
$\{N,~a_q,~a^\dagger_q\}$ will be reduced to the ordinary one as
the deformation-parameter $q$ approaches 1. The deformed
oscillator is related to the simple harmonic oscillator as follows
\cite{song}
\begin{equation}
a_q=a\sqrt{[N]_q\over N},~~~~a_q^\dagger=\sqrt{[N]_q\over
N}a^\dagger,~~~~N=a^\dagger a,\label{real}\end{equation} where we
have used the notation $[x]_q={q^x-q^{-x}\over q-q^{-1}}$, and the
ordinary boson commutation relations are
\begin{equation}[a,a^\dagger]=1,~~~~[N,a]=-a,~~~~[N,a^\dagger]=a^\dagger.\label{boson}\end{equation}
The representation of the deformed Weyl-Heisenberg algebra is
obtained by constructing the Fock space, \begin{equation}\mid
n\rangle_q={(a^\dagger_q)^n\over\sqrt{[n]!}}\mid
0\rangle,\label{rep}\end{equation} which satisfies
\begin{equation}a^\dagger_q\mid n\rangle_q=\sqrt{[n+1]_q}\mid
n+1\rangle_q,~~~~a_q\mid n\rangle_q=\sqrt{[n]_q}\mid
n-1\rangle_q,\label{aket}\end{equation} where $[n]_q!\equiv
[n]_q[n-1]_q\cdots[2]_q[1]_q$. Using the formalism of the boson
realization of the deformed algebra, it can be easily demonstrated
that the Fock basis of the deformed algebra is the same as the one
of the ordinary algebra, namely
\begin{equation}\mid
n\rangle_q={(a^\dagger_q)^n\over\sqrt{[n]!}}\mid 0\rangle=
{(a^\dagger)^n\over\sqrt{n!}}\mid 0\rangle=\mid
n\rangle.\label{ket}\end{equation}

In the following, Hawking's computation of black hole evaporation
will be briefly reviewed, then a radiation spectrum associated
with the $q$-deformed scalar particles will be derived by using
the above ordinary boson realization of the $q$-deformed algebra.

Consider, now, the background spacetime of a Schwarzschild black
hole appropriate to a collapsing spherical body. For simplicity,
we only consider a single massless scalar field $\phi$ which is
coupled to gravity minimally \cite{wald}. Near the past null
infinity \pn, the quantum field $\phi$ can be expanded as
\begin{equation}W_{\rm in}^{-1}\phi(x)W_{\rm in}=\sum\limits_\sigma(a_{\rm
in}(\sigma)\phi_{\rm in}(\sigma,x) +a^\dagger_{\rm
in}(\sigma)\overline{\phi}_{\rm
in}(\sigma,x)),\label{in}\end{equation} where the label ``in''
denotes the incoming process; $\{\fin\}$ is an orthonormal basis
of the one-particle Hilbert space \Hin which is constructed in
Minkowski spacetime, and $\overline{\phi}_{\rm in}$ represents the
complex conjugate function of $\fin$; $a^\dagger_{\rm in}$ and
$a_{\rm in}$ are boson creation and annihilation operators
corresponding to $\{\fin\}$; $W_{\rm in}$ denotes the isomorphism
map, $W_{\rm in}: {\mathcal L}\rightarrow{\mathcal L}_S({\mathcal
H}_{\rm in})$, which associates with each state in $\mathcal L$
the Minkowski spacetime state it ``looks like" in the past (here
${\cal L}$ denotes a Hilbert space, and ${\cal L}_S({\cal H})$
represents a symmetric Fock space, for the detail see
\cite{wald}); and the notation $\sigma$ is used to distinguish
different modes. Near the future null infinity \fn and the event
horizon of black hole, the field reads \begin{equation}W_{\rm
out}^{-1}\phi(x)W_{\rm out}=\sum\limits_\sigma(a_{\rm
out}(\sigma)\phi_{\rm out}(\sigma,x) +a^\dagger_{\rm
out}(\sigma)\overline{\phi}_{\rm
out}(\sigma,x)),\label{out}\end{equation} where the subindex
``out" denotes the outgoing process; $\{\fout\}$ is an orthonormal
basis of the one-particle Hilbert space \Hout that constructed in
flat spacetime, $a^\dagger_{\rm out}$ and $a_{\rm out}$ are boson
creation and annihilation operators corresponding to $\{\fout\}$;
$W_{\rm out}$ denotes the isomorphism map, $W_{\rm out}: {\mathcal
L}\rightarrow{\mathcal L}_S({\mathcal H}_{\rm out})$, which
associates with each state in $\mathcal L$ the Minkowski spacetime
state it ``looks like" in the future.

The next step is to extend $\fout$ to the whole background
spacetime. Therefore, near the past null infinity \pn, the form of
$\fout'$ (the prime is used to denote the extension) can be
expressed as the linear combination of $\fin$ and $\bfin$,
\begin{equation}\phi'_{\rm
out}(\sigma,x)=\sum\limits_{\sigma'}(A_{\sigma\sigma'} \phi_{\rm
in}(\sigma',x)+B_{\sigma\sigma'} \overline{\phi}_{\rm
in}(\sigma',x)),\label{out'}\end{equation} where
$A_{\sigma\sigma'}$ and $B_{\sigma\sigma'}$ are Bogoliubov
transformation coefficients, satisfying the following relations
\begin{equation}AA^\dagger-BB^\dagger=1,~~~~AB^T-BA^T=0.\label{bogoliubov}\end{equation}
Inversion of (\ref{out'}) leads to \begin{equation}\phi'_{\rm
in}(\sigma,x)=\sum\limits_{\sigma'}(A^\dagger_{\sigma\sigma'}
\phi_{\rm out}(\sigma',x)-B^T_{\sigma\sigma'} \overline{\phi}_{\rm
out}(\sigma',x)).\label{in'}\end{equation}

One of the important issues to consider is how the
characterization of the states of the field as ``in" states
compares with their characterization as ``out" states. This is
given by the $S$-matrix, \begin{equation}S=W_{\rm out}^{-1}W_{\rm
in}.\label{smatrix}\end{equation} Given any ``in" state
$\Psi\in{\mathcal L}_S({\mathcal H}_{\rm in})$ describing how the
state ``looks" at early times, the ``out" state $S\Psi\in{\mathcal
L}_S({\mathcal H}_{\rm out})$ describes how the state ``looks" at
late times. Hence, we have the following relation
\begin{equation}\mid\psi\rangle_{\rm out}=S\mid 0\rangle_{\rm
in}.\label{sket}\end{equation} This will tell us the spontaneous
creation of particles by the gravitational field of black hole.

Using (\ref{in})-(\ref{smatrix}), with the orthogonality and
completeness of the basis, one can obtain the relations between
the operators of the ``out" state and those of the ``in" state
\begin{equation}\begin{array}{l} S^{-1}a_{\rm
out}(\sigma)S=\sum\limits_{\sigma'}(a_{\rm
in}(\sigma')A^\dagger_{\sigma'\sigma}-a^\dagger_{\rm
in}(\sigma')B^\dagger_{\sigma'\sigma}),\\S^{-1}a^\dagger_{\rm
out}(\sigma)S=\sum\limits_{\sigma'}(a^\dagger_{\rm
in}(\sigma')A_{\sigma\sigma'}-a_{\rm
in}(\sigma')B_{\sigma\sigma'}).\end{array}\label{sa}\end{equation}
The expected number of particles that spontaneously created in the
mode $\sigma$ can be expressed as, \begin{equation}\langle
N(\sigma)\rangle=_{\rm out}\langle\psi\mid a^\dagger_{\rm
out}(\sigma)a_{\rm out}(\sigma)\mid\psi\rangle_{\rm
out}.\label{n1}\end{equation} By means of (\ref{sket}) and
(\ref{n1}), in addition with the Schwarzschild solution, the
distribution of particle number can be given \cite{hawking,wald}
\begin{equation}\begin{array}{ll}\langle N(\sigma)\rangle &=_{\rm in}\langle
0\mid S^{-1}a^\dagger_{\rm out}(\sigma)a_{\rm out}(\sigma)S\mid
0\rangle_{\rm
in}\\&\displaystyle=(BB^\dagger)_{\sigma\sigma}\\&\displaystyle={1\over
e^{\omega_\sigma/T_H}-1},\end{array}\label{n2}\end{equation} with
the form of standard thermal spectrum, which was first derived by
S. Hawking, where $T_H$ is the Hawking temperature
\begin{equation}T_H={\kappa\over 2\pi},\label{temp}\end{equation} with the surface gravity
$\kappa$ expressed as
\begin{equation}\kappa={1\over4GM},\label{surgra}\end{equation} where $G$ is the
Newton's gravity constant, and $M$ is the ADM energy of black
hole.

Above, the Hawking's theory of the evaporation of black hole was
briefly reviewed. We considered a background spacetime of a
Schwarzschild black hole from a collapsing spherical body, and
only a single massless scalar field with a minimal coupling to
gravity was taken into account for simplicity. The distribution of
radiation particle-number was proven to be of the form of a
perfect black-body spectrum according to the Hawking's theory. If
the spacetime noncommutativity is taken into account, it should be
expected that the radiation spectrum deviates from the standard
thermal-form. We have proposed to describe the noncommutativity by
means of the $q$-deformation scheme, {\it i.e.} we can mimic the
particles live on the noncommutative spacetime near black hole by
using the $q$-deformed particles (\ref{q}), from which a different
radiation spectrum can be derived. That means that we adopt an
effective viewpoint that all the noncommutative effects are
ascribed to the modification of the behavior of the radiation
field meanwhile the background metric is preserved as the usual
Schwarzschild metric. In the following, we will substitute the
$q$-deformed particles to the normal ones, and re-calculate the
black hole radiation spectrum in the framework of this toy model.

It has been proven that the basis of the Fock space would not be
deformed by the $q$-deformation, namely the representation of the
deformed algebra is the same as that of the normal algebra. That
is to say, the relationship between the ``out"-state and the
``in"-state will be stood as (\ref{n1}) under the $q$-deformation.
Hence, the expected number of $q$-particles in the $\sigma$-th
mode of black hole radiation under the consideration of the
spacetime noncommutativity will be
\begin{equation}\begin{array}{ll} \langle
N(\sigma)\rangle_q&=_{\rm out}\langle \psi\mid a_{\rm out}^\dagger
(\sigma)_q a_{\rm out}(\sigma)_q\mid\psi\rangle_{\rm
out}\hh&\displaystyle= _{\rm in}\langle 0\mid S^{-1}a_{\rm
out}^\dagger (\sigma)_q a_{\rm out}(\sigma)_q S\mid 0\rangle_{\rm
in}\hh&=_{\rm in}\langle 0\mid S^{-1}[N_{\rm out}(\sigma)]_q S\mid
0\rangle_{\rm in}\hh&=_{\rm in}\langle 0\mid S^{-1}{q^{N_{\rm
out}(\sigma)}-q^{-N_{\rm out}(\sigma)}\over q-q^{-1}} S\mid
0\rangle_{\rm in}\end{array}\label{nq}\end{equation} To evaluate
this expectation value, an approximation is used \cite{gas}:
\begin{equation}\langle N^k(\sigma)\rangle\simeq\langle
N(\sigma)\rangle^k,\label{nappr}\end{equation} and further we get
\begin{equation}\langle q^N\rangle\simeq q^{\langle
N\rangle}.\label{qappr}\end{equation} Therefore, the approximate
result can be obtained \begin{equation}\langle
N(\sigma)\rangle_q\simeq{q^{\langle N(\sigma)\rangle}-q^{-\langle
N(\sigma)\rangle}\over q-q^{-1}}=[\langle
N(\sigma)\rangle]_q.\label{nqappr}\end{equation} For convenience,
we re-express the deformation parameter as
\begin{equation}\eta=\ln q,\label{eta}\end{equation} Therefore, the expectation
value of the number of radiated particles in the state $\sigma$ of
the deformed system can finally be of the form
\begin{equation}\langle N(\sigma)\rangle_\eta={\sinh(\eta\langle
N(\sigma)\rangle)\over\sinh\eta}={\sinh({\eta\over
e^{\omega_\sigma/T_H}-1}
)\over\sinh\eta}.\label{neta}\end{equation} Next, let us calculate
the energy spectrum. The quantum numbers of every mode can be
labelled by $\omega lm$, namely $\sigma=(\omega lm)$. According to
the particle number distribution (\ref{neta}), the number of
particles with the angular momentum $l$ per unit time in the
frequency range $\omega$ to $\omega+d\omega$, passing out through
the surface of the sphere, can be got
\begin{equation}\rho(\eta,\omega)=(2l+1){\omega^2d\omega\over
2\pi^2}{\sigma_l(\omega)\sinh({\eta\over e^{\omega/T_H}-1}
)\over\sinh\eta},\label{num}\end{equation} where
$\sigma_l(\omega)$ is the grey-body factor, and $(2l+1)$ is the
degeneracy of the angular momentum. Thus the total outgoing energy
flux (luminosity) of the black hole is given by
\begin{equation}L_\eta={1\over 2\pi^2 \sinh\eta
}\sum\limits_{l=0}^\infty(2l+1)\int_0^\infty
d\omega\omega^3\sigma_l(\omega){\sinh({\eta\over e^{\omega/T_H}-1}
)}.\label{leta}\end{equation} This is just the total energy flux
expression of the massless scalar radiation field with
$q$-deformation that radiated by black hole under the
consideration of spacetime noncommutativity. As we set $\eta=0$,
the original Hawking radiation will be recovered, namely
\begin{equation}L_0={1\over 2\pi^2
}\sum\limits_{l=0}^\infty(2l+1)\int_0^\infty
d\omega\omega^3\sigma_l(\omega){1\over
e^{\omega/T_H}-1}.\label{l0}\end{equation} The difference between
(\ref{leta}) and (\ref{l0}) can be viewed as the consequence of
the spacetime noncommutativity. Furthermore, it implies that some
correlation effect among the radiation particles is introduced due
to the noncommutativity. This point seems helpful for
understanding the problem of information loss.

We now analyze the properties of the radiation energy-spectrum in
detail. Since the $s$-wave in the energy spectrum is dominant, in
addition that the gray-body factor of the $s$-wave is a constant,
we will only analyze the $s$-wave for convenience. The radiation
energy distribution of the $s$-wave can be expressed as
\begin{equation}\rho_s(\eta,\omega)=A{\omega^3\over 2\pi^2\sinh(\eta)}
\sinh({\eta\over e^{\omega/T_H}-1}),\label{rhos}\end{equation}
where $A$ is the $s$-wave gray-body factor, with the value event
horizon area of black hole, $A_H$. If the parameter $\eta$ is set
to be 0, the distribution (\ref{rhos}) will be reduced to the
standard black-body radiation distribution. However, if the
parameter $\eta$ deviates from 0, even if slightly, the shape of
the distribution function will be distorted evidently comparing to
the shape of the black-body radiation, especially in the infrared
region. Fig.1 shows the distributions as $\eta$ is taken to be 0,
0.8, 1.4, and 2, respectively (from top to bottom, near the wave
crests, in turn $\eta=0, 0.8, 1.4, 2$). We can see explicitly that
the infrared divergency occurs when $\eta$ deviates from 0. It
implies that the physics in infrared region will be influenced by
the ultraviolet modification that provided by the $q$-deformation.
This phenomenon seems to take on some UV/IR correspondence
principle. Moreover, it seems that some new physical mechanism
hides in the infrared region. A natural treatment within the
framework of our model is to introduce an infrared cut-off. It
should be pointed out that the infrared cut-off always be
necessary for the states count outside the black hole. In 't
Hooft's brick wall model, an infrared cut-off was imposed in the
WKB approximation for solving the number of states, $g(E)$
\cite{thooft}. For our prediction about the infrared cut-off, it
looks more natural. In addition, we should mention that, by
different considerations, Bekenstein and Mukhanov ever gave a
minimal frequency $\omega_0$, as the fundamental quanta of the
system of black hole, to characterize the quantum gravity of black
hole \cite{bekenstein} \begin{equation}\omega_0={\alpha\over
16\pi}{1\over 2GM},~~~~~~\alpha=4\ln2.\label{om0}\end{equation} Of
course, there still exists another possibility, that is our method
is unsuitable for the low frequency region. However, in any case,
an infrared cut-off is necessary.

\vskip.8cm
\begin{figure}
\begin{center}
\leavevmode \epsfbox{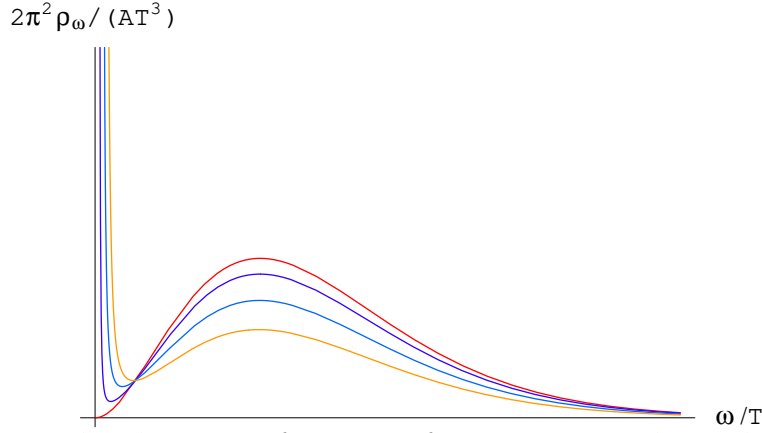} \caption[]{The energy distributions
vs frequency of $s$-wave as $\eta$ is taken to be 0, 0.8, 1.4, and
2, respectively (from top to bottom, near the wave crests, in turn
$\eta=0, 0.8, 1.4, 2$).}
\end{center}
\end{figure}

We now evaluate the position of the cut-off in frequencies. Since
the exact solution is difficult to get, we only give an
approximate evaluation. First, we have
\begin{equation}\rho_s\sim\left( \frac{\eta^3}{6\,{\left( -1 + e^x
\right) }^3} +
    \frac{\eta}{-1 + e^x} \right) \,x^3\label{rhosappr}\end{equation} where $x=\omega/T_H$,
    and we have expanded the sinh-function up to cubic term. For
    solving the extremum, we have
\begin{equation}\begin{array}{c}
\displaystyle 0={\partial\rho_s\over\partial x} \sim\frac{-\left(
\eta\,\left( 2\,{\left( -1 + e^x \right) }^2\,
         \left( 3 + e^x\,\left( -3 + x \right)  \right)  +
         \eta^2\,\left( 1 + e^x\,\left( -1 + x \right)
           \right)  \right) \,x^2 \right) }{2\,
    {\left( -1 + e^x \right) }^4}\\~
\\ \displaystyle\sim\frac{-\left( x\,\left( x^2 +
        2\,\left( -2 + x \right) \,x \right)  \right)
        }{2}\end{array}\label{diff}\end{equation}
In the second step, $e^x\simeq
    1+x$ is used. Then, the stationary points can be obtained
\begin{equation}\omega_{m1}=T_H \frac{1}{2}({2 - {\sqrt{2}}\,{\sqrt{2 -
\eta^2}}}),~~ \omega_{m2}=T_H \frac{1}{2}({2 +
{\sqrt{2}}\,{\sqrt{2 - \eta^2}}})\label{stationary}\end{equation}
In our opinion, the minimal frequency of the system, if it exists,
should be $\omega_{m1}$. In particular, for our interest, if we
identify the $\omega_{m1}$ with the Bekenstein's $\omega_0$, an
appropriate value of $\eta$ can be determined, that is
\begin{equation}\eta\sim 1.35\label{etav}\end{equation} Naively, we can view this value as
an important parameter to characterize the property of quantum
gravity of black hole in some sense. Fig.2 shows the situation of
$\eta=1.35$ contrasting to the standard case.

\vskip.8cm
\begin{figure}
\begin{center}
\leavevmode \epsfbox{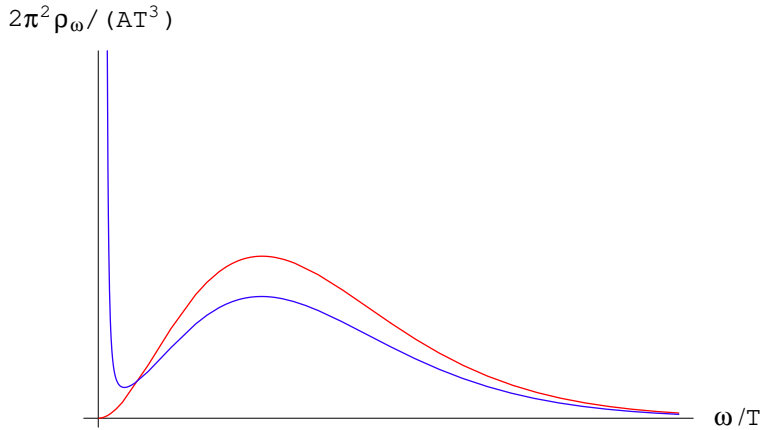} \caption[]{The spectrum with
$\eta=1.35$ comparing to the standard thermal spectrum.}
\end{center}
\end{figure}

Likewise, as an extension, for massless bosonic particles with
spin $s$, the total flux can be also derived by the same
$q$-deformation technique \begin{equation}{L}={g_s\over 2\pi^2
\sinh\eta }\sum\limits_{l=0}^\infty(2l+1)\int_0^\infty
d\omega\omega^3\sigma_{s,l}(\omega){\sinh({\eta\over
e^{\omega/T_H}-1} )},\label{sl}\end{equation} where $g_s$ is used
to denote the degeneracy of spin with the value 2 for nonzero spin
and 1 for zero spin, $\sigma_{s,l}(\omega)$ is the gray-body
factor of black hole.

Furthermore, we switch to Schwarzschild black hole in $D$
dimensions. The metric of such a black hole is
\begin{equation}ds^2=-(1-{r_H^{D-3}\over r^{D-3}})dt^2+(1-{r_H^{D-3}\over
r^{D-3}})^{-1}dr^2+r^2d\Omega_{D-2}^2.\label{metric}\end{equation}
The relation between the horizon radius and the black hole mass is
\begin{equation}r_H^{D-3}={16\pi G_D\over
(D-2)\Omega_{D-2}}M,\label{radius}\end{equation} where $G_D$ is
the Newton's gravity constant in $D$ dimensions and $M$ is the ADM
energy of black hole. The black hole has a Hawking temperature
\begin{equation}T_H={D-3\over 4\pi r_H}.\label{dtemp}\end{equation} The density of state of
the radiation can be written as
\begin{equation}g^{(D)}_sg^{(D)}_l{\Omega_{D-2}\omega^{D-2}d\omega\over(2\pi)^{D-1}},\label{density}\end{equation}
where $g^{(D)}_s$ and $g^{(D)}_l$ are used to denote the
degeneracies of spin and orbital angular momentum in $D$
dimensional spacetime respectively. We also give the explicit
expression of $\Omega_{D-2}$:
\begin{equation}\Omega_{D-2}={2(\sqrt{\pi})^{D-1}\over\Gamma[{D-1\over
2}]}.\label{domega}\end{equation} Then, the total flux of the
spontaneously created $q$-deformed particles by the gravitational
field of black hole in the $D$ dimensional spacetime, can be
obtained \begin{equation}{L}={\Omega_{D-2}g^{(D)}_s\over
(2\pi)^{D-1} \sinh\eta }\sum\limits_{l=0}^\infty
g^{(D)}_l\int_0^\infty
d\omega\omega^{D-1}\sigma_{s,l}(\omega){\sinh({\eta\over
e^{\omega/T_H}-1} )}.\label{dl}\end{equation}

In summary, a $q$-deformation prescription for introducing
spacetime noncommutative effects into the black hole evaporation
is proposed. We postulate a kind of $q$-deformed physical degrees
of freedom to characterize the effects come from spacetime
noncommutativity. The trick we use in this letter is that the
spacetime noncommutative effects are ascribed to the modification
of the behavior of the radiation field of black hole, and the
calculation is performed under the usual Schwarzschild metric.
Despite the suggestion of this toy model is highly speculative, it
is still an attempt to probe the effects of noncommutative quantum
gravity of black hole. A new spectrum of the black hole radiation
is obtained by means of the $q$-deformation scheme, which deviates
from the standard thermal spectrum evidently. It seems that some
correlation is introduced to the radiation system. The existence
of the infrared divergency implies that perhaps some unknown
physical mechanism plays a role in the infrared region. In the
framework of our model, an infrared cut-off is introduced. The
fact that the change of the behavior of the infrared part
originates from the ultraviolet modification embodies some UV/IR
correspondence principle.

\begin{acknowledgments}
The author would like to thank Zhe Chang, Tong Chen and Chao-Guang
Huang for useful discussions. The author also thanks Dmitri
Vassilevich for valuable comments. This work was supported in part
by the Natural Science Foundation of China (Grant No. 10375072).
\end{acknowledgments}


\end{document}